\documentclass[%
 reprint,
superscriptaddress,
 amsmath,amssymb,
 aps,
pra,
]{revtex4-2}

\usepackage{graphicx}
\usepackage{dcolumn}
\usepackage{bm}
\usepackage{amsmath}
\usepackage{amssymb}
\usepackage{color}
\usepackage{hyperref}

\begin{document}

\preprint{APS/123-QED}

\title{Realization of joint weak measurement in classical optics using optical beam shifts}

\author{Ritwik Dhara}
\affiliation{Department of Physical Sciences, Indian Institute of Science Education and Research (IISER) Kolkata, Mohanpur 741246, India.}
\author{Shyamal Guchhait}
\affiliation{Department of Bioengineering, University of Washington, Seattle, WA, USA}
\author{Meghna Sarkar}
\affiliation{Department of Physical Sciences, Indian Institute of Science Education and Research (IISER) Kolkata, Mohanpur 741246, India.}
\author{Swain Ashutosh}
\affiliation{School of Physics and Astronomy, University of Glasgow, Glasgow G12 8QQ, United Kingdom}
\author{Niladri Modak}
\email{niladri.modak@tuni.fi}
\affiliation{Photonics Laboratory, Physics Unit, Tampere University, Tampere 33720, Finland}
\author{Nirmalya Ghosh}
\email{nghosh@iiserkol.ac.in}
\affiliation{Department of Physical Sciences, Indian Institute of Science Education and Research (IISER) Kolkata, Mohanpur 741246, India.}
\affiliation{Center of Excellence in Space Sciences India, Indian Institute of Science Education and Research Kolkata, Mohanpur 741246, India}




 


\date{\today}

\begin{abstract}
Quantum weak measurements became extremely popular in classical optics to amplify small optical signals for fundamental interests and potential applications. Later, a more general extension, joint weak measurement has been proposed to extract weak value from a joint quantum measurement. However, the detection of joint weak value in the realm of classical optics remains less explored. Here, using the polarization-dependent longitudinal and transverse optical beam shift as a platform, we experimentally realize the quantum joint weak measurement in a classical optical setting. Polarization states are cleverly pre and post-selected, and different single and joint canonical position-momentum observables of the beam are experimentally extracted and subsequently analyzed for successful detection of complex joint weak value. We envision that this work will find usefulness for gaining fundamental insights on quantum measurements and to tackle analogous problems in optics.


\end{abstract}

\maketitle


\section{Introduction}
Unlike standard strong quantum measurement, `weak measurement' involves a weak coupling between the system and the pointer. Along with a clever postselection, it can give rise to anomalous enhancement of the outcome of the measurement, i.e., the `weak value' which can lie even outside the allowed eigenvalue spectrum of the measuring observable \cite{AAV,Duck_sudarshan,Hulet,hosten2008observation,kofman2012nonperturbative,resch2004extracting}. 
Such enhanced outcome is often termed as weak value amplification \cite{AAV,Duck_sudarshan,Hulet,hosten2008observation,kofman2012nonperturbative,resch2004extracting}. 
Although discovered in the context of quantum measurement, weak measurement and weak value amplification became extremely popular in classical optics, thanks to its inherent interferometric origin \cite{Duck_sudarshan,Hulet,hosten2008observation}. 
In fact, most of the experiments involving weak measurements have been performed in classical optical settings for small optical signal amplifications, such as, for the estimation of small phase\cite{xu2013phase}, quantifying small angular rotations \cite{magana2014amplification}, measuring ultrasmall time delays \cite{salazar2014measurement}, capturing tiny beam deflections \cite{dixon2009ultrasensitive} and so on \cite{zhong2024polarization,li2016application,he2021detection,qiu2016precisely}. More importantly, the implementation of weak measurement in optical domain yields improved understandings of different foundational aspects of quantum physics 
\cite{kocsis2011observing,lundeen2009experimental,goggin2011violation,lundeen2011direct}, as well as different optical phenomena, e.g, optical beam shifts \cite{modak2023longitudinal}
, spin-orbit interaction of light \cite{bliokh2015spin}, spin Hall effect of light \cite{hosten2008observation,bliokh2016spin} etc. 
\par
In 2004, Resch and Steinberg incorporated the idea of weak measurement in a more general scenario, a joint quantum measurement \cite{resch2004extracting}. The weak value of a joint measurement, namely `joint weak value' impacts the separable initial pointer states resulting in the postselection-dependent modifications in the joint canonical position and/or momentum variable (second moment) of the pointer, unlike just a shift (first moment) in the pointer variable in case of single weak measurement \cite{resch2004extracting}. As a result, this joint canonical variable encodes different important information about joint quantum processes \cite{yu2010joint,di2011strong,busch1984various,khanahmadi2021guessing,rehan2020experimental,ringbauer2014experimental}. Such joint weak measurements involving the product of two or more observables have provided useful insights into fundamental quantum mechanical processes, allowed characterization of the evolution of quantum systems \cite{ren2019efficient,hofmann2012weak}, enabled generation and characterization of multiparticle entanglement \cite{resch2004practical}, helped in studying quantum paradoxes \cite{lundeen2009experimental,yokota2009direct} and so on \cite{de2017single,strubi2013measuring}. Although joint weak measurements have been formulated and explored primarily in quantum systems, a classical optical implementation is capable of enriching our understanding of joint measurements in a relatively simple experimental platform \cite{hosten2008observation,kofman2012nonperturbative}. Moreover, just like usual optical weak measurements and weak value amplification, knowledge of joint weak values may also find useful metrological applications to quantify joint optical effects \cite{qin2023quality,modak2022interferometric,modak2023longitudinal} 
and simulate analogous quantum mechanical phenomena \cite{lundeen2005practical,raikisto2024joint}. 
Although efforts have been made recently to access optical settings \cite{puentes2012weak,kobayashi2012extracting,yokota2009direct,lundeen2009experimental,qin2023quality,strubi2013measuring,zhu2022joint}
, the classical optical realization of joint weak measurement with separable input pointer state ,i.e., the original proposal of Resch and Steinberg \cite{resch2004extracting}, still remains unexplored.
\par
In this paper, we experimentally realize the single particle joint weak measurement in a classical optical setting -- in a total internal reflection (TIR) of a Gaussian beam. A Gaussian beam when total internally reflecting from a dielectric interface experiences polarization-dependent Goos-Hänchen (GH) and Imbert-Fedorov (IF) beam deflection \cite{bliokh2013goos}. Importantly, the simultaneous appearance of the GH and IF shifts \cite{goswami2014simultaneous} enables such a system to act as a natural classical optical platform that mimics a quantum joint weak measurement. Different single and joint canonical position-momentum observables of the reflected beam are experimentally extracted and subsequently analyzed for successful detection of complex joint weak value. We anticipated that our study will open up a relatively simple route to study and simulate a number of classical optical analogue of several quantum mechanical events and quantum measurements using various degrees of freedom of classical light beam \cite{lundeen2005practical,raikisto2024joint,kobayashi2012extracting,yokota2009direct,lundeen2009experimental,qin2023quality,strubi2013measuring}. 

\section{Theoretical framework for observing JWV}
\begin{figure}[!h]
    \centering
    \includegraphics[width=1\linewidth]{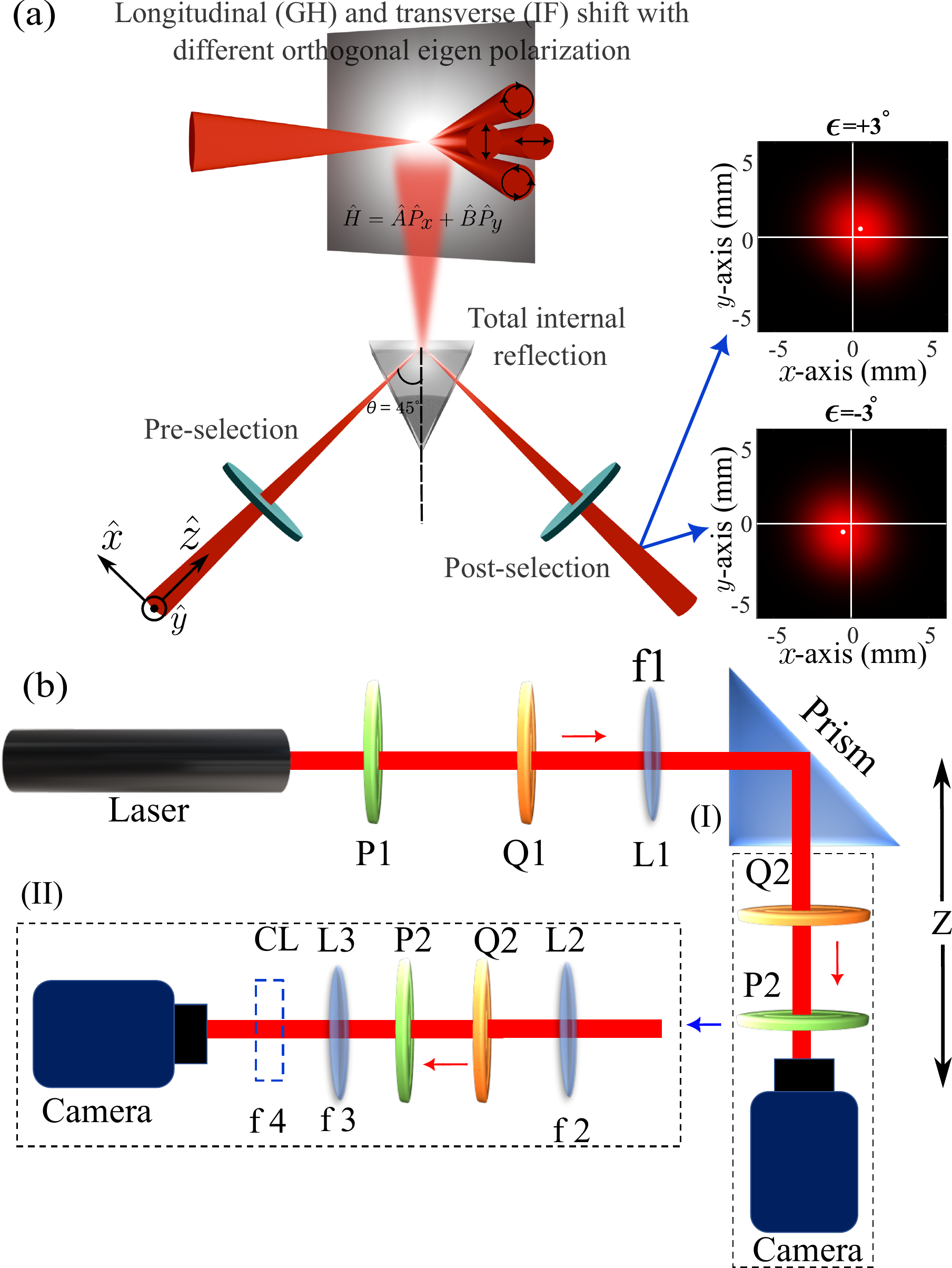}
    \caption{Schematic illustration of joint weak measurement in GH and IF shifts, and the sketch of the experimental setup to detect complex JWV. (a) Simulatneous presence of polarization-depndent GH and IF shifts mimic an ideal joint weak measurement scenario. The eigenpolarizations of GH shift are horizontal and vertical linear polarization, whereas IF shift possesses opposite handed circular (elliptical) polarizations \cite{gotte2014eigenpolarizations}. The beam experiencing TIR is judiciously pre and postselected to enhance single and joint weak value. The enhancement of both the shifts is further demonstrated by the opposite-digonal beam deflection for near orthogonal postselections at $\epsilon=\pm3^{o}$ in the simulated beam profiles. (b) Schematic experimental setup for (I) separate determination of real and imaginary part of the weak value of GH and IF shift. (II) Required modification of the setup in the reflected path for the detection of the joint pointer variables $\left\langle XY\right\rangle_{fi}$ and $\left\langle XP_{y}\right\rangle_{fi}$. }
    \label{Fig:reim}
\end{figure}

We consider a polarized Gaussian beam of wavelength $\lambda$ undergoing total internal reflection (TIR) from a glass-air interface at an angle of incidence $\theta$. The transverse profile of the beam can be expressed as a function of local Cartesian coordinate $x-y$, where $x$ is the coordinate in the plane of incidence, while $y$ is perpendicular to this plane and the associated operators are $\hat{X}$ and $\hat{Y}$ respectively (see Fig. \ref{Fig:reim}(a)). The corresponding momentum are $p_{x}$ and $p_{y}$ with their associated operators $\hat{P}_x$ and $\hat{P}_y$. The two types of beam shifts, i.e., GH and IF shift, represented by the Artmann operators $\hat{A}$ and $\hat{B}$, correspond to the spatial beam deflection in longitudinal (in $x$-direction) and transverse (i.e., in $y$-direction) direction respectively \cite{bliokh2013goos},
\begin{equation}
A =  -\dfrac{1}{k}\begin{pmatrix}
    \dfrac{\partial \delta_p}{\partial\theta} & 0\\
    0 & \dfrac{\partial \delta_s}{\partial\theta}
\end{pmatrix};\ B = \dfrac{i\cot\theta}{k}
\begin{pmatrix}
    0 & 1+e^{i\delta}\\
    -(1+e^{-i\delta}) & 0
\end{pmatrix}
\label{eqn:GH}
\end{equation}
where $e^{i\delta_{p}}$ and $e^{i\delta_{s}}$ are the Fresnel's reflection coefficients with
$\delta = \delta_p - \delta_s$, $k = 2\pi/\lambda$ (see Appendix \ref{app:shift_matrices}). 
Both these shifts are very small in magnitude when compared to input beam dimension and thus mimic an ideal weak measurement scenario \cite{hosten2008observation}.  
Weak value amplification, therefore has played a key role in experimentally observing these beam shifts for answering many fundamental questions related to beam shifts \cite{dennis2012analogy} and to open up novel metrological applications \cite{qin2023quality,zhong2023dual}. Interestingly, GH and IF shifts always appear simultaneously in case of the reflection or transmission of a realistic light beam owing to their non-separability \cite{modak2023longitudinal}. Additionally, for a TIR, both the Artmann operators $\hat{A}$ and $\hat{B}$ for GH and IF shifts are Hermitian and non-commuting \cite{dennis2012analogy}, mimicking a joint weak measurement scenario proposed by Resch and Steinberg (see Fig. \ref{Fig:reim}(a)) \cite{resch2004extracting}. Therefore, the corresponding Hamiltonian takes the form $\hat{H}=\hat{A}\hat{P}_x+\hat{B}\hat{P}_y$ representing momentum-domain evolution of the reflected beam (see Fig. \ref{Fig:reim}(a)). Therefore, we can use the expressions (given in \cite{resch2004extracting}) to extract complex joint weak value $(AB+BA)_{W}$  of $\hat{A}$ and $\hat{B}$ in our TIR system.
 \begin{equation}
 \label{eqn:Re}
     Re\left\langle\dfrac{AB+BA}{2}\right\rangle_{W} = 2\left(\frac{2\pi}{\lambda}\right)^2\left\langle XY\right\rangle_{fi}-Re(\left\langle A\right\rangle_{W}^{*}\left\langle B\right\rangle_{W})
 \end{equation}
 \begin{equation}
   \begin{split}
 \label{eqn:Im}
     Im\left\langle\dfrac{AB+BA}{2}\right\rangle_{W} &= 4 \left(2w_{y0}\frac{2\pi}{\lambda}\right)^2\left\langle XP_y\right\rangle_{fi}\\&-Im(\left\langle A\right\rangle_{W}^{*}\left\langle B\right\rangle_{W})  
 \end{split}  
 \end{equation}
 where $\left\langle A\right\rangle_{W}$ ($=\frac{\langle\psi_f|\hat{A}|\psi_i\rangle}{\langle\psi_f|\psi_i\rangle}$ where $|\psi_i\rangle$ is the preselected state, and $|\psi_f\rangle$ is the postselected state), $\left\langle B\right\rangle_{W}$ are the single weak value of observables $\hat{A}$ and $\hat{B}$ respectively \cite{resch2004extracting}. $\left\langle XY\right\rangle_{fi}$, $\left\langle XP_y\right\rangle_{fi}$ are the expectation values of joint observables $XY$ and $XP_y$ after successful postselection \cite{resch2004extracting},
 $w_{y0}$ is the radius of the pointer distribution in $y$-direction. 

\par
\begin{figure}[h!]
    \centering
\includegraphics[width=1\linewidth]{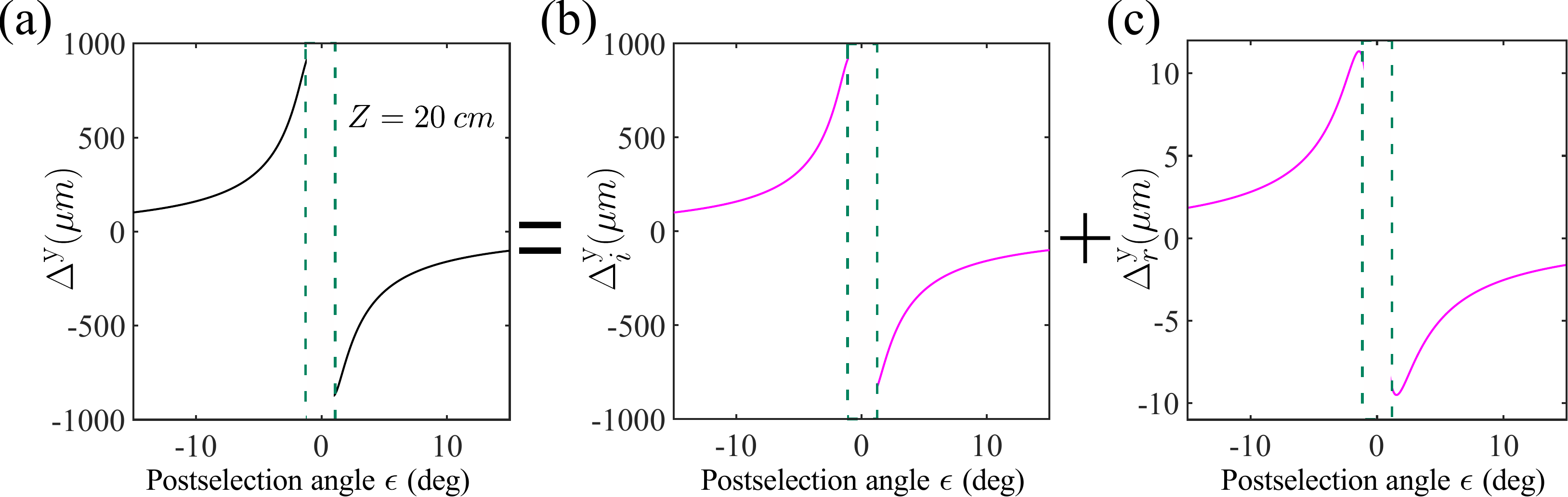}
    \caption{Simulated centroid shift of a Gaussian beam undergoing TIR at a propagation distance $Z=20\ cm$ considering wavelength $\lambda = 633\ nm$, Rayleigh range $Z_0 = 608.19\ \mu m$, the range of postselection angle $\epsilon$ is chosen following our experimental situation. The total shift can be expressed as a combination of imaginary (momentum-domain/angular) shift and real (position-domain/spatial) shift both of which show weak value amplification beavior ($\sim \cot{\epsilon}$) \cite{goswami2014simultaneous}. At $\epsilon\rightarrow 0$ (region shown by green dashed line), weak value does not follow the beam shift trajectory and also the beam fails to maintain its Gaussian nature.}
    \label{Fig:20_IF_sim}
\end{figure}

 Now we discuss the strategy to extract JWV from our experiment. Note that, the real and imaginary parts of the single weak value, are associated with spatial and angular shift of the reflected beam respectively \cite{aiello2008role}. The total shift of the reflected beam at a distance $Z$ (see Fig. \ref{Fig:20_IF_sim}) can be expressed as \cite{toppel2013goos} 
\begin{equation}
  \label{eqn:Prop}
     \Delta^{y} =\Delta^{y}_i +\Delta^{y}_r = \dfrac{Z}{Z_{0}}Im(\left\langle B\right\rangle _{W}) + Re(\left\langle B\right\rangle_{W})
\end{equation}
where $Z_{0}$ is the Rayleigh range \cite{toppel2013goos}. Here $\Delta^{y}$ corresponds to the shift in the $y$-direction and $\Delta^{y}_i, \Delta^{y}_r$ are related to imaginary and the real part of the the weak value $\langle B \rangle_W$ (see Fig. \ref{Fig:20_IF_sim}) causing momentum and space-domain beam shifts respectively. Similar equation is valid for the shift along the $x$-direction and can be expressed in terms of $\Delta^{x}$ and $\left\langle A\right\rangle_{W}$. Note that in Fig. \ref{Fig:20_IF_sim}(a), the total shift is calcuated from the simulated experiment (see Appendix \ref{app: Simulation}). On the other hand, Fig. \ref{Fig:20_IF_sim}(b),(c) are extracted from theoretical values of the two terms in the right hand side of Eq. \eqref{eqn:Prop}, respectively.
\\
Note that to extract the real and imaginary part of $\langle B\rangle_W$ (Eq. \eqref{eqn:Prop}), we consider two different propagation distances $Z1$ and $Z2$ for the measurement of the shifts $\Delta_{1}^{y}$ and $\Delta_{2}^{y}$ respectively.
 After solving those two equations, we get 
 \begin{equation}
     \label{eqn:im}
     Im(\left\langle B\right\rangle _{W}) = \dfrac{Z_0(\Delta^y_2 - \Delta^y_1)}{Z2-Z1}
 \end{equation}
 \begin{equation}
     \label{eqn:re}
     Re(\left\langle B\right\rangle _{W})=\Delta^y_2 - \dfrac{Z2}{Z_0}Im(\left\langle B\right\rangle _{W})
 \end{equation}
 With these equations, we can calculate the terms $Re(\left\langle A\right\rangle_{W}^{*}\left\langle B\right\rangle_{W})$ and $Im(\left\langle A\right\rangle_{W}^{*}\left\langle B\right\rangle_{W})$ present in Eq. \eqref{eqn:Re}, \eqref{eqn:Im} which have the form,
\begin{equation}
    \begin{split}
    \label{Re}
      Re(\left\langle A\right\rangle_{W}^{*}\left\langle B\right\rangle_{W}) & = Re(\left\langle A\right\rangle_{W})Re(\left\langle B\right\rangle_{W})\\ &+Im(\left\langle A\right\rangle_{W})Im(\left\langle B\right\rangle_{W})    
    \end{split}
\end{equation}
\begin{equation}    
\begin{split}
\label{Im}
   Im(\left\langle A\right\rangle_{W}^{*}\left\langle B\right\rangle_{W}) &=  Re(\left\langle A\right\rangle_{W})Im(\left\langle B\right\rangle_{W})\\
    & -Re(\left\langle B\right\rangle_{W})Im(\left\langle A\right\rangle_{W}). 
\end{split}
\end{equation}  
Next, $\left\langle XY\right\rangle_{fi}$ and $\left\langle XP_y\right\rangle_{fi}$ are extracted from the image of postselected beam. Therefore, with the knowledge of all these parameters, we can completely extract the complex joint weak value using Eq. \eqref{eqn:Re},\eqref{eqn:Im}. Now we turn to the demonstration of the experimental detection of joint weak value.
\section{Experimental setup}
The schematic experimental setup is presented in Fig. \ref{Fig:reim} (b). A He-Ne Laser (HNL050L, Thorlabs, USA) of wavelength, $\lambda = 633\ nm$ having beam-waist $\xi_y= 2.73 \ mm$ is used to seed the system. A rotatable linear polarizer P1 (LPVISE100-A, Thorlabs, USA) and a quarter waveplate Q1 (WPQ10M-633, Thorlabs, USA) are used to preselect the polarization state of the incident beam. To further enhance the magnitude of beam shifts, a biconvex lens L1 of focal length $75\ mm$ is employed to focus the beam at the desired interface of the glass-air interface of the prism \cite{pal2019experimental,jayaswal2013weak,jayaswal2014observation}. The radius of the beam at the interface is $w_0$ is $11.035\ \mu m$ which sets the Rayleigh range $Z_0$ at $608.19\ \mu m$. In our experiment, $w_0$ and $w_{y0}$ (see Eq. \eqref{eqn:Im}) are equal since the two dimensional projection of our pointer is circular in nature. Another set of quarter waveplate (Q2) and linear polarizer (P2) mounted on motorized precision rotation mount (PRM1/MZ8, Thorlabs, USA) postselects the polarization state of the reflected beam. The reflected beam is recorded by a CCD camera (Andor iKon-M, $24\ \mu m$ pixel dimension) at two different distances $Z1$ ($12\ cm$) and $Z2$ ($20\ cm$) from the point of reflection (see Fig. \ref{Fig:fin_both_GH}). 


To detect joint pointer variable $\left\langle XY\right\rangle_{fi}$, the postselection part (I) is replaced by (II) (see Fig. \ref{Fig:reim}) where a conventional $4-f$ system consisting of two spherical lenses L2 (focal length $f2 = 50\ mm$) and L3 (focal length $f3 = 400\ mm$) is introduced. The $4-f$ system images the point of interaction on the camera placed at the front focal plane of L3 as shown in Fig. \ref{Fig:reim} \cite{thekkadath2016direct}. Note that this combination of lenses also magnifies the spot size of the beam $10.66$ times at the detection point making the centroid detection more accurate (see Appendix \ref{app:amplification_factor} ). To detect $\left\langle XP_{y}\right\rangle_{fi}$, in addition to the $4-f$ system, we introduced a cylindrical lens CL (focal length $f4 = 50\ mm$) in the front focal plane of L3. CL is oriented in such a way that it gives the Fourier transform of only $y$-direction, without any effect on the $x$-direction \cite{thekkadath2016direct}. 

\section{Results and Discussions}
To experimentally observe the joint weak values, we choose the pre and postselection in such a way that the corresponding pointer observables are enhanced, leading to greater experimental ease in observing them. $\begin{pmatrix}1 & 1\end{pmatrix}^T$ state, i.e., $+45^{o}$ linear polarization is chosen as the preselection and the subsequent near orthogonal elliptical postselection $\left\lvert\psi_{f}\right\rangle$ =
 $\begin{pmatrix}
     1-i(\sin 2\epsilon + \cos 2\epsilon) & -1-i(\sin 2\epsilon - \cos 2\epsilon)
\end{pmatrix}^{T}$ enhances both the GH and IF shift simultaneously \cite{goswami2014simultaneous}. $\epsilon$ is the postselection angle denoting the offset to the exact orthogonal pre and postselection. This pre and postselection combination enhances both the real and imaginary parts of $\left\langle A\right\rangle_{W}$ and $\left\langle B\right\rangle_{W}$ (see Appendix \ref{app:Pol_states}) which are necessary for the computation of JWV as mentioned in Eq. \eqref{eqn:Re}, \eqref{eqn:Im} \cite{resch2004extracting}.\\

First, we experimentally measure the shift of the centroid of the beam in both $x$ and $y$-directions, $\Delta^{x}$ and $\Delta^{y}$ respectively. 
\begin{figure}[!h]
    \centering
    \includegraphics[width=1\linewidth]{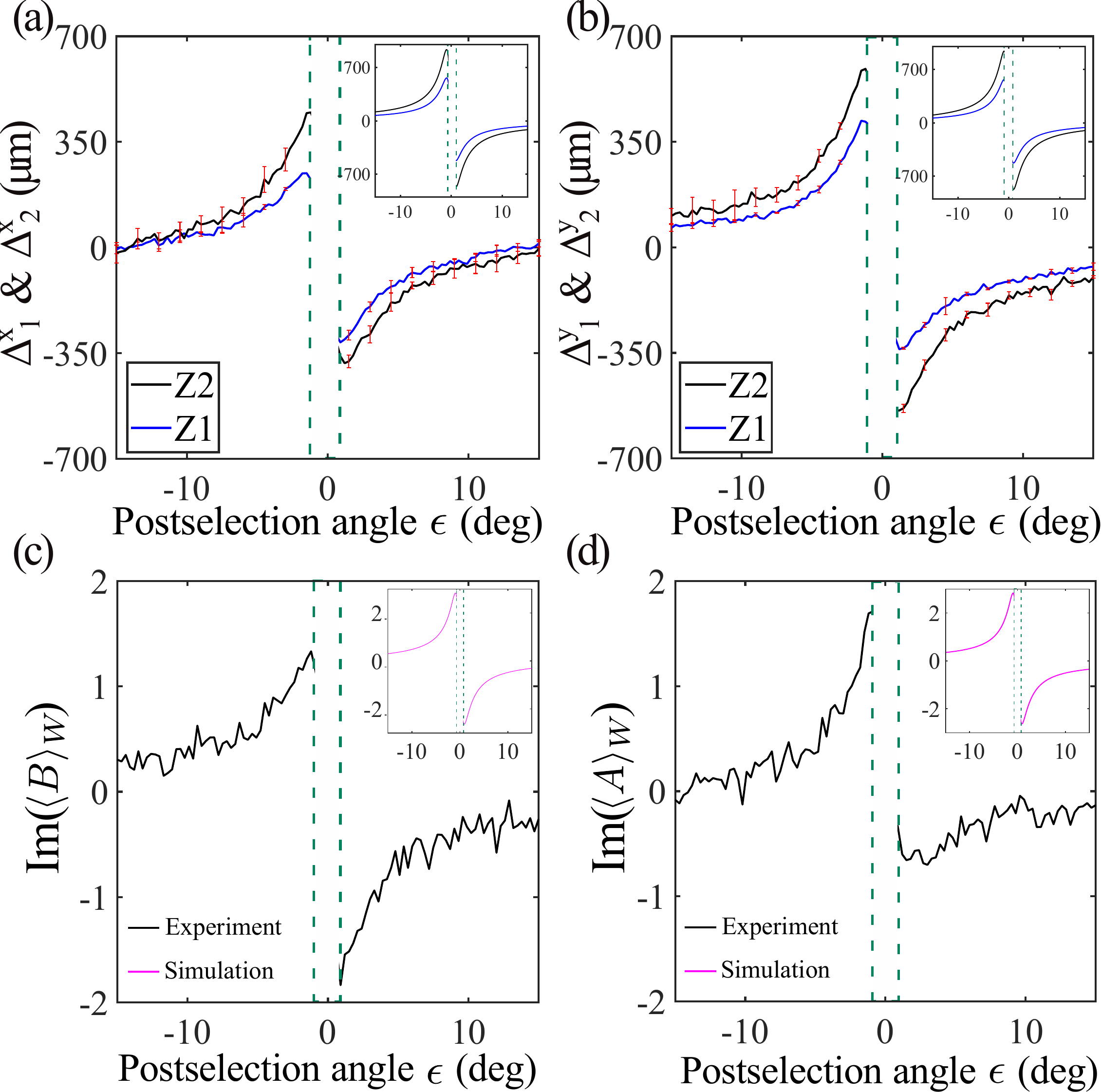}
    \caption{Extraction of single weak values for GH and IF shifts. 
    (a),(b) Experimentally detected GH shift $\Delta^{x}$ and IF shift $\Delta^{y}$ of the centroid of the beam respectively with changing postselection angle $\epsilon$, at two different distances $Z1=12cm$ (blue solid line) and $Z2=20cm$ (black solid line) at the incident angle $\theta = 45^\circ$. The insets show corresponding simulation results (see Appendix \ref{app: Simulation}). Error bars represent statistical fluctuations.
    (c),(d) show the extracted (using Eq. \eqref{eqn:im}) imaginary parts of $\left\langle A\right\rangle _{W}$ and $\left\langle B\right\rangle _{W}$ respectively with changing postselection angle $\epsilon$. Insets show the corresponding simulation results. The region where the centroid of the beam is undetectable is marked by green dashed lines.
    }
    \label{Fig:fin_both_GH}
\end{figure}
The xperimentally obtained and simulated variations of $\Delta^{x}$ and $\Delta^{y}$ with changing postselection angle $\epsilon$ are shown in Fig. \ref{Fig:fin_both_GH}(a),(b) for two detection positions $Z1=12cm$ and $Z2=20cm$. $\Delta^{x}$ and $\Delta^{y}$ show the expected weak value amplification 
behavior \cite{pal2019experimental} and agrees with the simulation. Note that the centroid of the beam cannot be detected around the zero offset angle, i.e., postselection angle $\epsilon=0$ as the Gaussian nature of the beam profile cannot be maintained \cite{gotte2013limits} and the intensity of the beam is almost comparable to the external noise level that results in error-prone centroid detection. These regions are marked by the green dashed lines in the figures (see Fig. \ref{Fig:fin_both_GH}). Next, $\Delta^{x}$ and $\Delta^{y}$ are used to separately extract the imaginary and real parts of the weak value using Eq. \eqref{eqn:im},\eqref{eqn:re}. The variations of the extracted imaginary parts of the weak value $\left\langle A\right\rangle _{W}$ and $\left\langle B\right\rangle _{W}$ with the change in the postselection angle $\epsilon$ are in good agreement with the corresponding prediction by simulation (see Fig. \ref{Fig:fin_both_GH}(c),(d)).
\par
Real weak value can be implicitly calculated using Eq. \eqref{eqn:re}. However, as apparent from Eq. \eqref{eqn:re} a small deviation of the experimental imaginary weak value from the theoretical one may lead to a large overestimation of the real part of $\left<B\right>_W$ ($\left<A\right>_W$). This, in turn, will lead to unwanted contribution of real weak values in JWV further (see Eq. \eqref{eqn:Re},\eqref{eqn:Im}).
In the Appendix \ref{app: Re_WV_detection}, we highlight that how a small difference of the estimated imaginary weak value leads to a large error in the real weak value. As also apparent from Fig. \ref{Fig:fin_both_GH}, some mismatch between the simulation and the experiment results can be observed. The asymmetric nature of the GH shift in the experimental plots (see Fig. \ref{Fig:fin_both_GH}) 
appears due to the finite pointer width in $x$-direction (in the plane of incidence) which causes deviation in the angle of incidence from $45^\circ$. As the critical angle of glass-air interface is very close to $45^\circ$, some part of the beam does not experience TIR. Additionally, these shift measurements are very prone to errors in the beam parameters and the detection geometry \cite{modak2022tunable}. Slight changes of these parameters may lead to the observed deviation in the quantitative values.
\begin{figure}[h!]
   \centering
  \includegraphics[width=1\columnwidth]{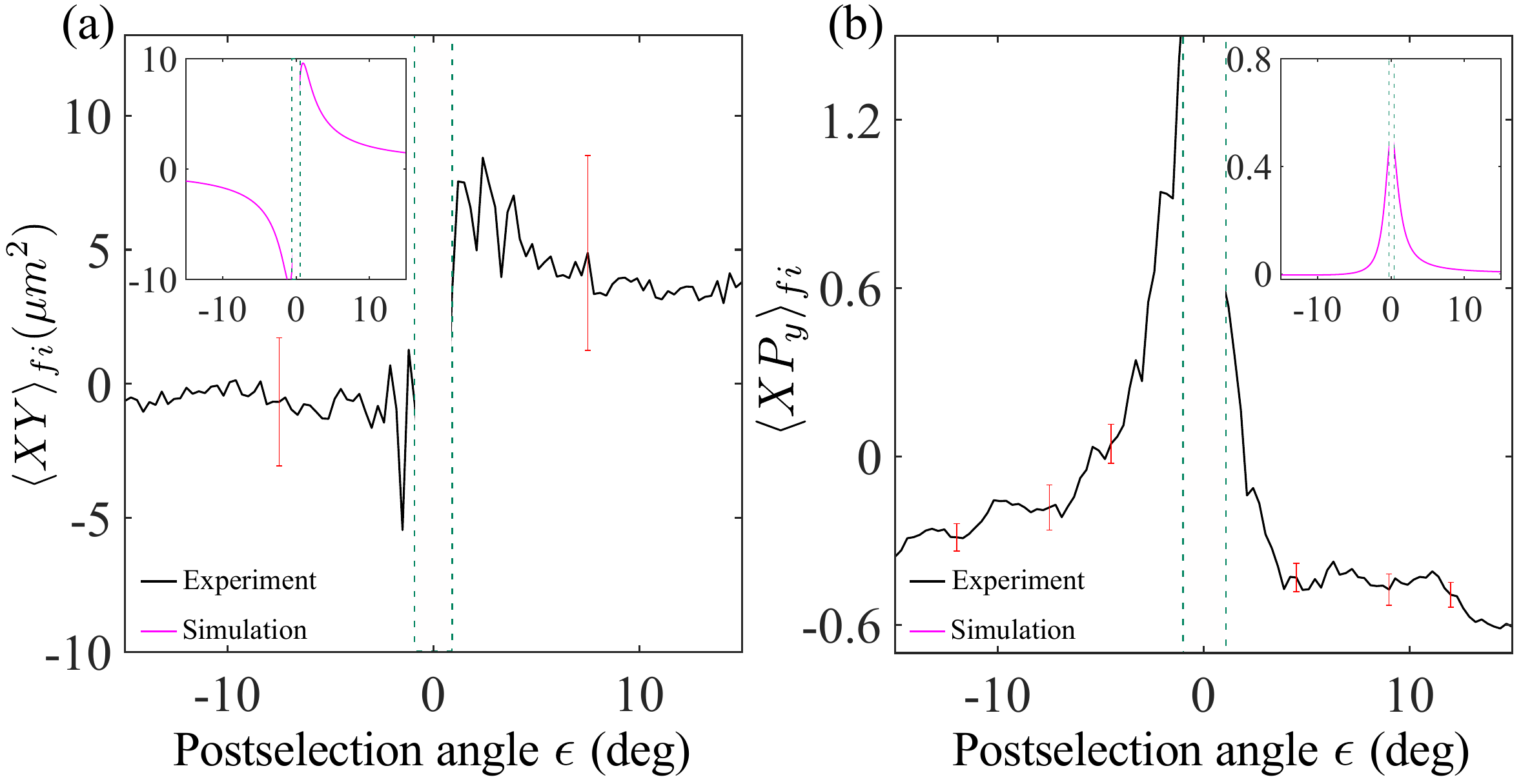}
    \caption{Experimentally observed variation of (a) $\left\langle XY\right\rangle_{fi}$ and (b) $\left\langle XP_{y}\right\rangle_{fi}$ with changing postselection angle $\epsilon$. Insets illustrate corresponding simulated variation (see Appendix \ref{app: Simulation}). Error bars represent statistical flactuations.}
    \label{Fig:XY_at_1mml}
\end{figure}
\par
Having experimentally determined the real and imaginary parts of $\left\langle A\right\rangle _{W}$ and $\left\langle B\right\rangle _{W}$, we now proceed to determine the joint pointer variables $\left\langle XY\right\rangle_{fi}$ and $\left\langle XP_{y}\right\rangle_{fi}$. In Fig. \ref{Fig:XY_at_1mml}, we show the variations of experimentally detected joint variables with changing postselection angle $\epsilon$ which are in significant agreement with the simulation predictions. 
Details of the simulations are discussed in Appendix \ref{app: Simulation}. Again, the asymmetric nature is reflected in both $\left\langle XY\right\rangle_{fi}$ and $\left\langle XP_{y}\right\rangle_{fi}$ (Fig. \ref{Fig:XY_at_1mml}), since both of these variables also involve GH shift.
\par
Considering all the experimentally detected parameters present in Eq. \eqref{eqn:Re},\eqref{eqn:Im} we calculate JWV. It can be noted from Fig. \ref{Fig:20_IF_sim} that in our experimental regime,  the contribution of the real weak value to the total shift is negligible as compared to that of the imaginary weak value. So we also neglect the first term of Eq. \eqref{Re} involving the product of two real weak values while calculating the experimental JWV further. The variations of the real and imaginary parts of JWV with change of postselection angle $\epsilon$ show excellent agreement with the exact theoretical predictions by Eq. \eqref{eqn:Re},\eqref{eqn:Im}. The slight mismatch in the imaginary part originated from the mismatch in the detection of $\left<XP_y\right>_{fi}$ as shown in Fig. \ref{Fig:XY_at_1mml}(b) which again arises from slight error prone estimations of experimental parameters, like spot size, detection distance, pixel size of the camera etc.
\begin{figure}
    \centering
    \includegraphics[width=1.0\columnwidth]{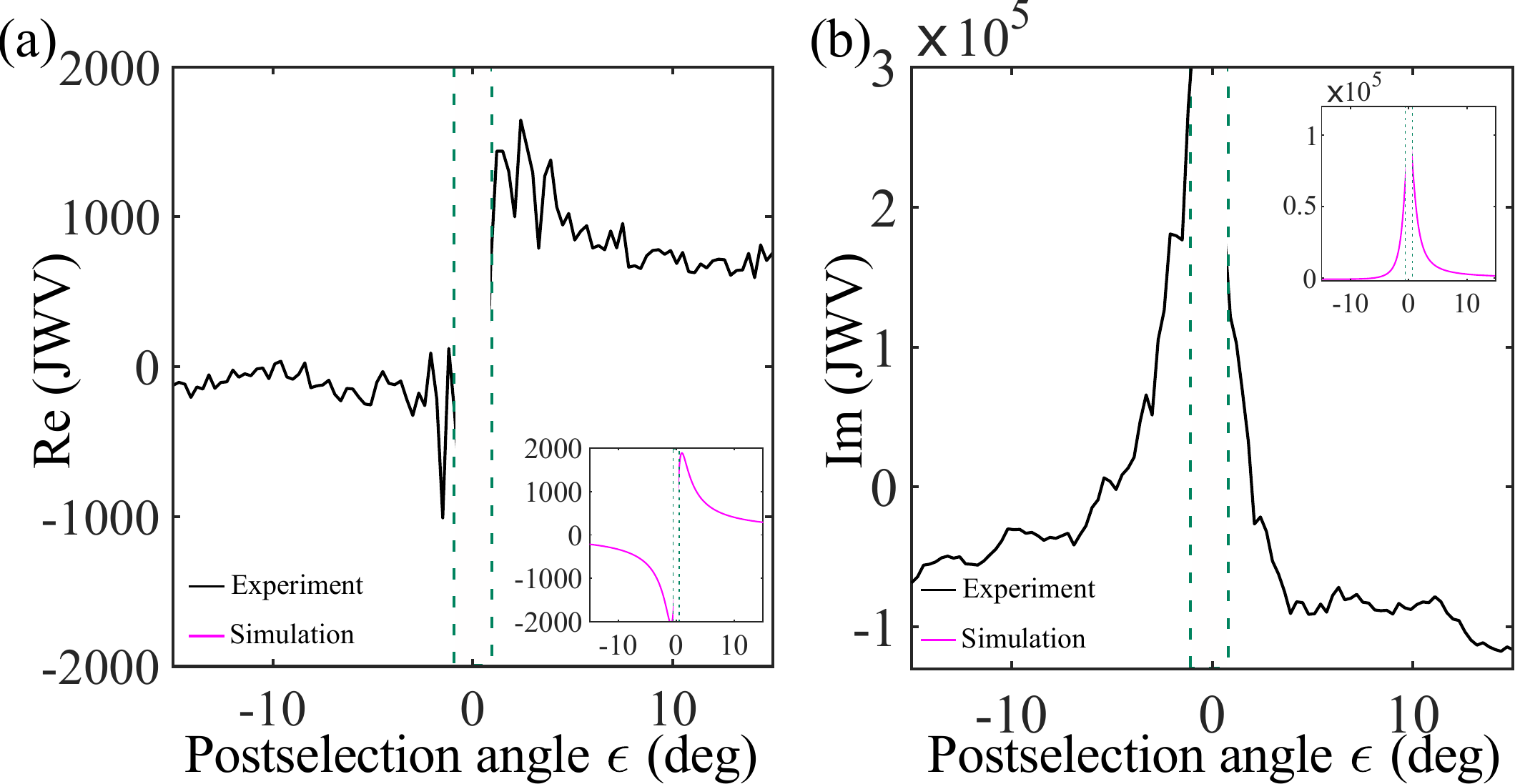}
    \caption{Variation of retrieved (a) real and (b) imaginary parts of the (JWV) from experiment. Insets show corresponding theoretical predictions (Eq. \eqref{eqn:Re},\eqref{eqn:Im}).}
    \label{Fig:JWV}
\end{figure}\\
This way, we detect the complex joint weak value using optical beam shifts in a Gaussian beam undergoing TIR. Note that with the presented protocol, several other variants of the optical beam shifts and multiple other degrees of freedom of light may now be explored to realize joint weak values \cite{bliokh2013goos,bliokh2016spin}. However, caution must be exercised while directly applying the JWV formulism to other variants of the beam shifts involving structured light and structured polarization. First of all, this approach is only applicable to Hermitian operators. In contrast, for example, in partial reflection, the beam shift operators are generally non-Hermitian in nature \cite{dennis2012analogy}. Moreover, in the case of the beam shift in complex structured light, e.g. Laguerre-Gaussian modes \cite{rubinsztein2016roadmap,zhu2022joint}
, the position and the momentum-domain pointer deflections corresponding to the beam shifts may get coupled in a rather complicated fashion giving rise to the mutual contributions between beam shifts along the longitudinal and transverse directions \cite{kobayashi2012extracting,bliokh2009goos}



\section{Summary} 
In summary, we have implemented a joint weak measurement scheme and detected the JWV of two observables in a classical optical setting using an optical beam shift platform. A Gaussian beam undergoing TIR from a dielectric interface experiences polarization-dependent GH and IF shifts in longitudinal and transverse spatial directions respectively, mimicking an ideal join weak measurement scenario as originally proposed by Resch and Steinberg \cite{resch2004extracting}. Different single and joint position-momentum canonical observables are experimentally detected and subsequently analyzed to extract complex JWV of the beam shift operators. This work opens up a new avenue to explore the physics of joint measurement in the realm of simpler classical optical experiments. Besides fundamental interests to gain deeper understanding of a number of analogous non-trivial optical phenomena \cite{lundeen2005practical,raikisto2024joint,kobayashi2012extracting,yokota2009direct,lundeen2009experimental,qin2023quality,strubi2013measuring}, such joint weak measurements using optical beam shifts may find useful practical applications in metrology and sensing using novel experimental metrics associated with the joint weak values of appropriately chosen degrees of freedom of the light beam \cite{strubi2013measuring,qin2023quality,fang2016ultra}.  

\section*{Acknowledgement}
The authors acknowledge the Indian Institute of Science Education and Research, Kolkata for the funding and facilities. RD acknowledges Ministry of Education, Govt. of India for PMRF fellowship and research grant. NM acknowledges the support of the Photonics Research and Innovation Flagship (PREIN - Decision 346511).
\appendix
\section{\MakeUppercase{Beam shift matrices in case of TIR}}
\label{app:shift_matrices}
In total internal reflection, the GH and IF shift matrices have the following form \cite{bliokh2013goos} 
\begin{equation}
\begin{split}
A = \dfrac{i}{k}
\begin{pmatrix}
    \dfrac{\partial ln r_p}{\partial\theta} & 0\\
    0 & \dfrac{\partial ln r_s}{\partial\theta}
\end{pmatrix}
= -\dfrac{1}{k}\begin{pmatrix}
    \dfrac{\partial \delta_p}{\partial\theta} & 0\\
    0 & \dfrac{\partial \delta_s}{\partial\theta}
\end{pmatrix};
\label{eqn:GH}
\end{split}
\end{equation}

\begin{equation}
\begin{split}
  B &= \dfrac{i\cot\theta}{k}
\begin{pmatrix}
    0 & 1+\dfrac{r_{p}}{r_{s}}\\
    -1-\dfrac{r_{s}}{r_{p}} & 0
\end{pmatrix}\\
&= \dfrac{i\cot\theta}{k}
\begin{pmatrix}
    0 & 1+e^{i\delta}\\
    -1-e^{-i\delta} & 0
\end{pmatrix}\\
&= \dfrac{\cot\theta}{k}[-(1+\cos\delta)\sigma_{y}-\sin\delta\sigma_{x}]
\label{eqn:IF}  
\end{split}
\end{equation}

$\delta_p$, $\delta_s$ are the phase part of Fresnel reflection coefficients($r_p$ and $r_s$) in TIR where $r_p = e^{i\delta_p}$, $r_s = e^{i\delta_s}$, $\delta=\delta_p - \delta_s$, $\theta$ is angle of incidence and $k=2\pi/\lambda$ and $\epsilon$ is post selection parameter. Here $\sigma_x$ and $\sigma_y$ are the Pauli matrices.
$\delta_p$ and $\delta_s$ has the form,
\begin{equation}
    \delta_p = -2\arctan\left( \frac{n\sqrt{n^2\sin{\theta}^2-1}}{\cos{\theta}}
 \right)
\end{equation}
\begin{equation}
     \delta_s = -2\arctan\left( \frac{\sqrt{n^2\sin{\theta}^2-1}}{n\cos{\theta}}
 \right)
\end{equation}
where $n=n_1/n_2$, $n_1$ is refractive index of glass and $n_2$ is refractive index of air.

\section{SIMULATION}
\label{app: Simulation}
\par For the simulation, we use MATLAB as the computational tool. A Gaussian beam is created with a spatial grid of $512\times512$ pixels, with each pixel having dimensions of $24\ \mu m$, which matches to that of the experimental setup. The Gaussian beam is generated using the equation
\begin{equation}
    \centering
    E \propto e^{k\left(iz-\frac{x^2+y^2}{2(z_0+iz)}\right)}
\end{equation}
where $x$ and $y$ represent the transverse coordinates, while $z_0$, $z$, and $k$ denote the Rayleigh range, the distance from the beam waist, and the wave vector, respectively.
The polarization of the Gaussian beam is pre-selected by $(1/r_p \quad 1/r_s)^T$. $r_p$ and $r_s$ are the Fresnel reflection coefficients in TIR. The medium considered is a total internal reflection (TIR) medium, with refractive indices $n_1 = 1.5$ and $n_2 = 1.0$. The beam propagates from the glass (higher refractive index) into the air (lower refractive index) and strikes the interface at an angle $\theta=45^\circ$. The Jones matrix of the surface, which defines the geometry of the reflection, is expressed by
\begin{equation}
    \centering
    M = \begin{pmatrix}
        r_p\left(1-i\frac{x}{z_0+iz}\frac{\delta \ln{r_p}}{\delta \theta}\right) & i\frac{y}{z_0+iz}(r_p+r_s)\cot{\theta} \\
        -i\frac{y}{z_0+iz}(r_p+r_s)\cot{\theta} & r_s\left(1-i\frac{x}{z_0+iz}\frac{\delta \ln{r_s}}{\delta \theta}\right)
    \end{pmatrix}
\end{equation}
Subsequent to the reflection, the beam is projected into a post-selected state defined by $(1 - i(\sin2\epsilon+\cos2\epsilon) \quad -1 - i(\sin2\epsilon-\cos2\epsilon))^T$, which is near an orthogonal state, with a small offset $\epsilon$ introduced to deviate slightly from the exact orthogonal configuration. In the simulation, the calculation of the beam centroid and first order weak values was performed analogously to the experimental procedure. 


\par Now we move on to discuss the extraction of joint observables from the simulation. Note that when we perform the simulation for detecting joint observables $\langle XY \rangle_{fi}$ and $\langle XP_y \rangle_{fi}$, instead of taking the distribution of the beam at the point of reflection, we choose to propagate the beam at a very small distance, as at the point of interaction the beam width is very small, i.e., $\sim 1$ pixel in our spatial grid, which results in errors in the subsequent calculations of canonical position momentum observables. We perform the simulation for several distances ranging $1.6-1.7\ mm$ from the reflection point, where we get the considerable amount of pixels in the beam image. Throughout this regime, the simulation results agree quite well with the corresponding experimentally obtained variations. We present the simulation result in Fig. \ref{Fig:XY_at_1mml}(a) for the distance $1.68\ mm$ from the point of interaction where the simulation result is the closest to the corresponding experimental one. A similar procedure is followed during the simulation of $\langle XP_y \rangle_{fi}$, i.e., first we allow the beam to propagate at some distance after reflection, then we perform the fast Fourier transformation (FFT) to the $y$-coordinate, with the $x$-coordinate remaining unchanged of the image. $\langle XY \rangle_{fi}$ and $\langle XP_y \rangle_{fi}$ are calculated from polarization postselected images following Appendix \ref{app:Image_analysis}.

\section{\MakeUppercase{Calculation of the amplification factor of the beam} }
\label{app:amplification_factor}

To measure the amplification factor used in the detection part of $\left < XY \right>_{fi}$ and $\left< XP_y \right>_{fi}$ we have used a single slit with known slit width of $0.5 \ mm$, at the back focal plane of L2 (Same point at which the beam will interact in air-glass interface for TIR). Then we used the same conventional $4-f$ system as used in Fig. \ref{Fig:reim}(b)(II) in the main text and project the image of the slit width on the front focal plane of L3. Then, we took an image of the slit width which we got $5.33\ mm$ so in our setup amplification factor is $10.66$. We follow the same technique with an additional cylindrical lens during the amplification factor measurement of $\left < XP_y \right>_{fi}$ term. 

\begin{figure}[h!]
    \centering
     \includegraphics[width=1.0\columnwidth]{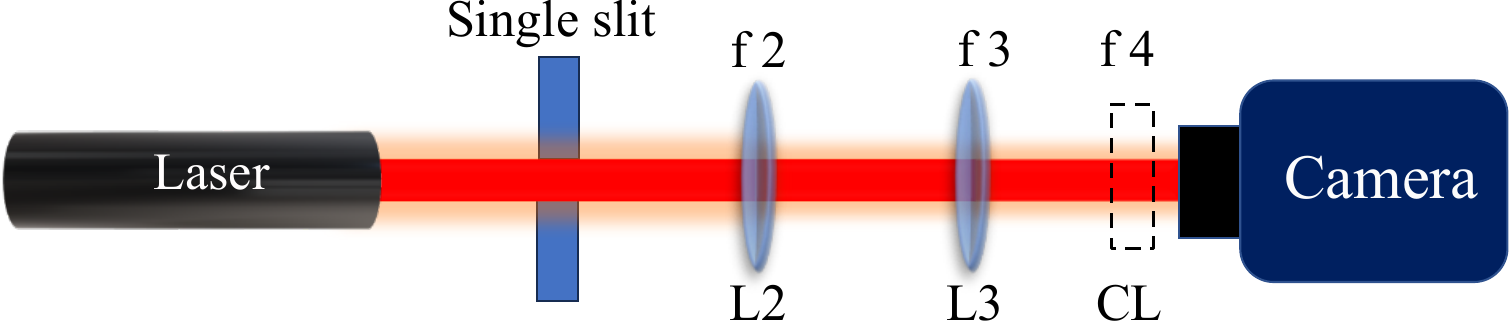}
    \caption{Experimental setup for the detection of amplification factor of the beam for the measurement of $\langle XY \rangle_{fi}$ and $\langle XP_y \rangle_{fi}$. L1 and L2 are the two spherical lenses having focal lengths $f2 = 50\ mm$  and $f3 = 400\ mm$. CL is the cylindrical lense having focal length $f4 = 50\ mm$.}
    \label{fig:enter-label}
\end{figure}

\section{\MakeUppercase{ Choice of pre and post-selected polarization state}}
\label{app:Pol_states}

Unlike plane waves, real Gaussian beams carry a finite Gaussian distribution of wave vectors around the central wave vector which leads to
the shifts in the centroid of its transverse profile when reflected or refracted from any interface \cite{bliokh2013goos}. The in-plane (Longitudinal) Goos-H\"anchen (GH) shift originates due to angular gradient of the Fresnel coefficients associated with the change of angle of incidence for the non-central wave vectors and the out of the plane (Transverse) Imbert-Fedorov (IF) shifts originates from the spin orbit interaction of light \cite{bliokh2013goos}.
Both types of shifts can be expressed in operator form mentioned in the main text(Operator $A$ and $B$). The eigenvalues of the shift matrices ($A$ and $B$) provide the maximal centroid shift of the beam when the
incident polarization coincides with the corresponding eigenvectors \cite{toppel2013goos}. In this work, we have chosen $\begin{pmatrix}
    1/r_p & 1/r_s
\end{pmatrix}^T$ which becomes $\begin{pmatrix}
    1 & 1
\end{pmatrix}^T$ after interaction with Fresnel-Jones matrix $\begin{pmatrix}
    r_p & 0 \\ 0 & r_s
\end{pmatrix}$. The pre-selection state $\left\lvert\psi_{i}\right\rangle$ =  $\begin{pmatrix}
    1 & 1
\end{pmatrix}^T$ can be written as a superposition of the eigenstates of GH shift matrices and as well as IF shift matrices below the Brewster region. The nearly orthogonal post-selection state  is, $\left\lvert\psi_{f}\right\rangle$ =$\begin{pmatrix}
     1-i(\sin 2\epsilon + \cos 2\epsilon) & -1-i(\sin 2\epsilon - \cos 2\epsilon)
 \end{pmatrix}^{T}$ which amplifies both the real and imaginary parts of both shifts.
 
\begin{equation}
\begin{split}
 \left\langle A \right\rangle _W &= 
\frac{\left\langle \psi_{f}\lvert A \rvert \psi_{i} \right\rangle}{\left\langle \psi_{f} \lvert \psi_{i}\right\rangle} \\
&= -\frac{1}{2k} \left[\left(\frac{\partial\delta_p}{\partial\theta}+\frac{\partial\delta_s}{\partial\theta}\right) + \left( \frac{\partial\delta_p}{\delta\theta} - \frac{\partial\delta_s}{\partial\theta}\right)\cot{2\epsilon} \right]\\
&+\frac{i}{2k\sin{2\epsilon}} \left[ \left( \frac{\partial\delta_p}{\partial\theta} - \frac{\partial\delta_s}{\partial\theta} \right) \right]
\label{eq:weakvalueA}
\end{split}
\end{equation}

\begin{equation}
\begin{split}
   \left\langle B \right\rangle _W &= 
 \frac{\left\langle \psi_{f}\lvert B \rvert \psi_{i} \right\rangle}{\left\langle \psi_{f} \lvert \psi_{i}\right\rangle}\\
 &=\frac{\cot{\theta}}{k}\left[  \frac{1+\cos{\delta}}{\sin{2\epsilon}} -\sin{\delta} \right]\\
 &+ \frac{i\cot{\theta}}{k}\left[ \left( 1+ \cos{\delta} \right) \cot{2\epsilon}\right]
 \label{eq:weakvalueB} 
\end{split}
\end{equation}
 From Eq.\eqref{eq:weakvalueA},\eqref{eq:weakvalueB} we can write $\left\langle A \right\rangle_W$, $\left\langle B \right\rangle_W$ as 

\begin{equation}
 \left\langle A \right\rangle _W = 
\frac{\left\langle \psi_{f}\lvert A \rvert \psi_{i} \right\rangle}{\left\langle \psi_{f} \lvert \psi_{i}\right\rangle} = Re(\left\langle A \right\rangle_W) + iIm(\left\langle A \right\rangle_W)
\label{S5}
\end{equation}

\begin{equation}
    \left\langle B \right\rangle _W = 
\frac{\left\langle \psi_{f}\lvert B \rvert \psi_{i} \right\rangle}{\left\langle \psi_{f} \lvert \psi_{i}\right\rangle} = Re(\left\langle B \right\rangle_W) + iIm(\left\langle B \right\rangle_W)
\label{S6}
\end{equation}
Where
\begin{equation}
  Re(\left\langle A \right\rangle_W) =
  -\frac{1}{2k} \left[\left(\frac{\partial\delta_p}{\partial\theta}+\frac{\partial\delta_s}{\partial\theta}\right) + \left( \frac{\partial\delta_p}{\delta\theta} - \frac{\partial\delta_s}{\partial\theta}\right)\cot{2\epsilon} \right]
  \label{ReA}
\end{equation}

\begin{equation}
    Im(\left\langle A \right\rangle_W) =
    \frac{1}{2k\sin{2\epsilon}} \left[ \left( \frac{\partial\delta_p}{\partial\theta} - \frac{\partial\delta_s}{\partial\theta} \right) \right]
    \label{ImA}
\end{equation}

\begin{equation}
    Re(\left\langle B \right\rangle_W) =
 \frac{\cot{\theta}}{k}\left[  \frac{1+\cos{\delta}}{\sin{2\epsilon}} -\sin{\delta} \right]
 \label{ReB}
\end{equation}

\begin{equation}
    Im(\left\langle B \right\rangle_W) =
    \frac{\cot{\theta}}{k}\left[ \left( 1+ \cos{\delta} \right) \cot{2\epsilon}\right]
    \label{ImB}
\end{equation}
All four parts of Eq. \eqref{S5},\eqref{S6} have an amplification term $\cot{2\epsilon}$ or $\dfrac{1}{\sin{2\epsilon}}$, which amplifies as $\epsilon$ goes near to zero. It is possible to detect both real and imaginary parts of GH and IF shift using these pre-post selection combination.

\section{\MakeUppercase{Beam's spatial profile analysis}}
\label{app:Image_analysis}
We use MATLAB to analyze the image. The pixel configuration of our camera is $512 \times 512$ and pixel size is $24\mu m$. Therefore, in MATLAB, the image is read as a $512\times512$ matrix, where each element represents a pixel and the value of that element represents the pixel's intensity. The method to estimate $\left\langle X \right\rangle$, $\left\langle Y \right\rangle$, $\left\langle XY \right\rangle$, $\left\langle XP_y \right\rangle$ from an experimentally obtained image of the light beam is covered in the following \cite{modak2023longitudinal}.
\begin{enumerate}
\item We take the background signal blocking the Total Internally Reflected beam to nullify the contribution of stray lights. This image is taken at the beginning of each experiment,
and we do not change the experimental setting once this image is taken. Next, we remove the reference beam's intensity element-by-element from the recorded image to ensure that the image used for analysis is free of any evidence of a continuous background noise source.
\item We do a coordinate transformation and take the centroid of the unshifted beam $(x_0,y_0)$ (centroid of the image
taken while the pre and post-selection are same) as the origin of our Cartesian system. We transform the coordinate system of the
CCD sensor to the centroid of the unshifted image to define our working Cartesian system. Next, we use that coordinate system to represent any particular image. This process is executed by
modifying $x$ and $y$ of any image with $x-x_0$ and $y-y_0$ respectively for $x$ and $y$ coordinate of
each pixel. Thus, we obtain the actual value of $\left\langle X \right\rangle$ , $\left\langle Y \right\rangle$, $\left\langle XY \right\rangle$, and $\left\langle XP_y \right\rangle$ with respect to the unshifted
beam.
\item We crop all the images before starting the analysis with reference to the centroid of the
unshifted beam ($x_0$, $y_0$) (centroid of the image). To find the centroid, we carry out an element-wise discrete sum of the coordinates ($x$ or $y$) over
the intensity profile of the image. The centroid in the $x$ and $y$ direction and their correlation
can be defined respectively as follows
\begin{equation}
   \left\langle X \right\rangle = \frac{\sum(x-x_0)I}{\sum I} ,   \left\langle Y \right\rangle = \frac{\sum(y-y_0)I}{\sum I}
\end{equation}
\begin{equation}
\left\langle XY \right\rangle = \frac{\sum(x-x_0)I\sum(y-y_0)}{\sum I}    
\end{equation}
where $I$ is the intensity of the corresponding image. We follow the same procedure to estimate $\langle XP_y \rangle$ as followed in the estimation of $\langle XY \rangle$. The only difference is that, in case of $\langle XP_y \rangle$ the shift in $y$ direction is the measurement is $p_y$ instead of $y$. 
\end{enumerate}
In the theoretical calculation same process is followed and all the parameter is calculated in the same way from the simulated post selected beam.

\section{\MakeUppercase{ Experimental extraction of real part of weak value for GH and IF shifts}}
\label{app: Re_WV_detection}

\begin{figure}[h!]
    \centering
    \includegraphics[width=1\columnwidth]{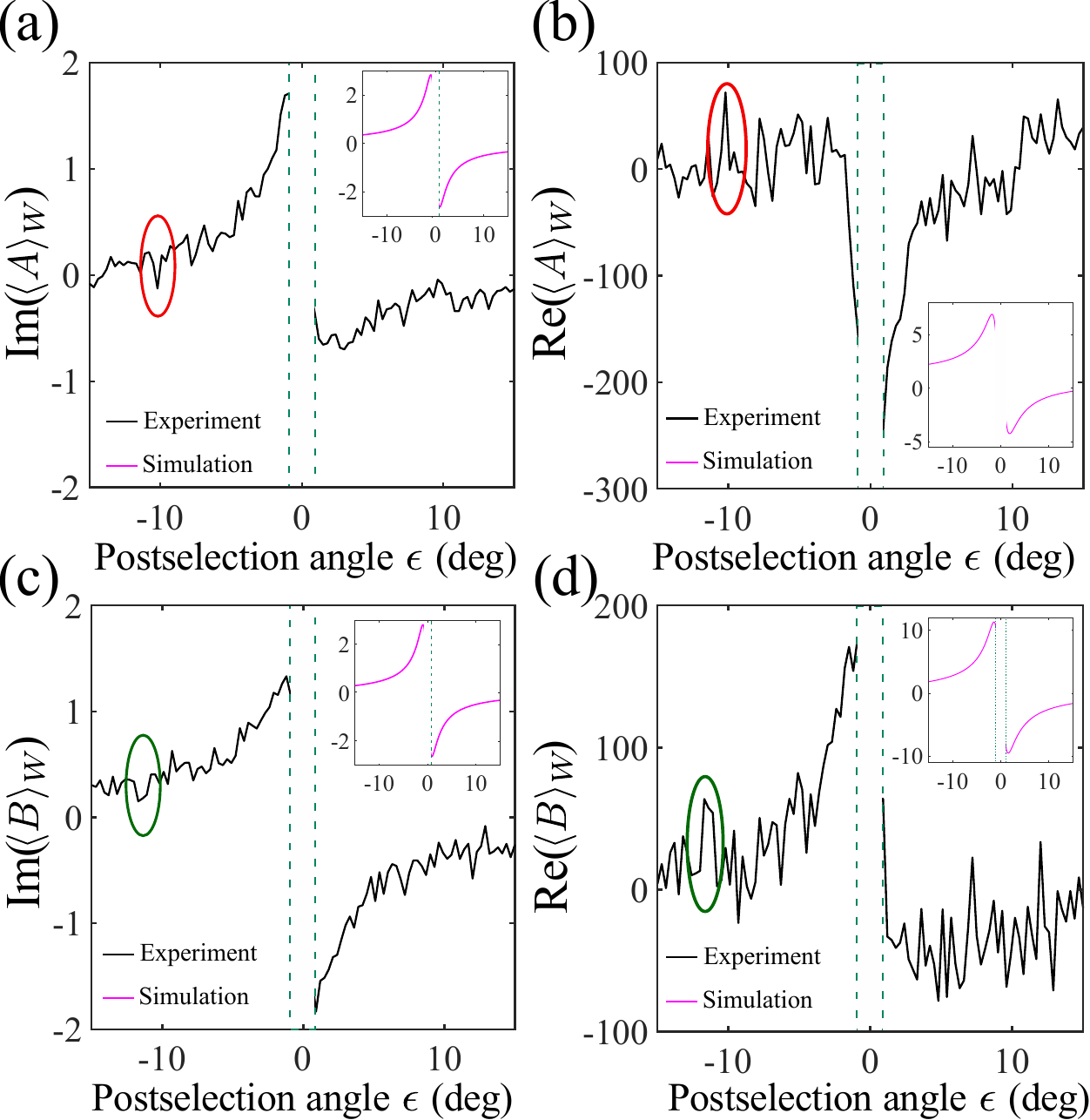}
    \caption{Experimentally retrieved real and imaginary parts of weak values for both the GH (top panel) and IF (bottom panel) shifts. (a),(b) show the comparison between experimentally determined values and the simulation of the imaginary and real parts of $\left\langle A\right\rangle _{W}$ respectively; (c),(d) demonstrate the contrast between values obtained by experimentation and the simulated imaginary and real components of $\left\langle B\right\rangle _{W}$ respectively.The purpose of this figure is to demonstrate how a small deviation in imaginary weak value leads to a huge error in the estimation of real weak value contradicting to the fact that the actual real weak value has a lower order of magnitude.} 
    \label{fig:both_reim}
\end{figure}

As discussed in the main text, imaginary weak valued are used to detect real weak values. A small deviation in imaginary weak value leads to a huge error in real weak value. These are shown in Fig. \ref{fig:both_reim}. The red and green circled region shows how a small fluctuation of the experimental data point in the imaginary part leads to a huge error in the real part for both the shifts. Although it is clear from the Fig. \ref{Fig:20_IF_sim} in the main text that the contribution of the real part of the shift in total shift is around $100$ times smaller that the imaginary shift. As a result, we have dropped the $2nd$ order term consisting of the real parts of the weak values from the Joint weak value experssion (Eq. \ref{eqn:Re}).       

\newpage
\nocite{*}
\bibliography{apssamp}

\end{document}